\documentstyle[12pt]{article}
\newcommand{\doublespace}{
    \renewcommand{\baselinestretch}{1.6}\large\normalsize}

\newcommand{\be}{\begin{equation}}
\newcommand{\ee}{\end{equation}}
\newcommand{\ba}{\begin{eqnarray}}
\newcommand{\ea}{\end{eqnarray}}
\newcommand{\ket}[1]{| {#1} \rangle}
\newcommand{\bra}[1]{\langle {#1} |}
\newcommand{\ave}[1]{\langle {#1} \rangle}

\newcommand{\btau}{\mbox{\boldmath $\tau$}}
\def\={\;=\;}
\def\+{\;+\;}
\def\psibar{\bar\psi}
\def\pbp{\langle\psibar\psi\rangle}
\def\psidagger{\psi^\dagger}
\def\pdp{\langle\psidagger\psi\rangle}

\def\roughly#1{\mathrel{\raise.3ex\hbox{$#1$\kern-.75em%
\lower1ex\hbox{$\sim$}}}}

\def\gsim{\roughly>}

\def\idp{\int\frac{d^3p}{(2\pi)^3}}

\begin{document}
\begin{titlepage}
\pagestyle{empty}
\vspace{1.0in}
\begin{flushright}
SUNY-NTG-96-7
\end{flushright}
\begin{flushright}
August 1996
\end{flushright}
\vspace{1.0in}
\begin{center}
\doublespace
\begin{large}
{\bf THE PROBLEM OF MATTER STABILITY IN THE NAMBU--JONA-LASINIO MODEL}\\
\end{large}
\vskip 1.0in
Michael Buballa\footnote{present address: Inst. f. Kernphysik, TH Darmstadt, 
Schlo{\ss}gartenstr. 9, 64289 Darmstadt, Germany}\\
{\small
{\it Department of Physics, State University of New York,\\
Stony Brook, New York 11794, U.S.A.}}\\
\end{center}
\vspace{2cm}

\begin{abstract}
We reinvestigate the conditions for stable matter solutions in the
Nambu--Jona-Lasinio (NJL) model. In mean field approximation the NJL model can
be regarded as an extension of the Walecka mean field model to include negative 
energy fermion states. While this extension is necessary to allow for a chiral
phase transition, it was found some time ago that at the same time it destroys
the wanted saturation properties of the Walecka model. We reformulate this
problem in terms of the thermodynamic potential and find that there is indeed
a connection between these two features. We show that the minimum of the
thermodynamic potential which corresponds to stable nuclear matter in the
Walecka model is shifted from a finite to zero effective fermion mass in the
chiral NJL model. This shift is closely related to the chiral phase transition.
Under certain conditions the shifted minima may still lead to stable matter
solutions but only in the chirally restored phase. We discuss a possible
interpretation of these solutions as a schematic bag model description. 
\end{abstract}
\end{titlepage}

\doublespace

\section{Introduction} 

\setcounter{equation}{0}

One of the major goals in intermediate energy physics is the understanding of
strongly interacting matter as a function of temperature and density and the
transition from the hadronic phase to the quark-gluon plasma. Unfortunately
nature provides us only with few points in the $T$-$\rho$ plane, namely the 
vacuum ($T=\rho=0$) and nuclear matter ($T=0$, $\rho = \rho_o =$ .17 fm$^{-3}$)
and much experimental and theoretical effort is necessary to obtain 
information about other regions. A major problem is that in general it is not
possible to prepare a static system of a given temperature and density. So
most points in the $T$-$\rho$ plane can only be reached for a short instant of
time, e.g. on a trajectory of expanding matter produced in a heavy ion
collision. For zero density  and finite temperature the most reliable 
existing ``data'' therefore come from lattice calculations, indicating that
both, the chiral phase transition and the confinement-deconfinement phase 
transition take place at the same temperature $T_c \simeq 150$ MeV \cite{Ka}. 
However, because of conceptual difficulties \cite{LK95p}, there are no firm 
lattice results at finite densities so far. 

In this situation simplified models may give us at least some ideas about 
hadronic matter at finite temperatures and densities which is necessary for 
the interpretation of the experimental data. Recent results on the lattice
\cite{KK95} and in random matrix models \cite{Ve95} indicate that the finite
temperature phase transition behaves like a mean-field transition at least
until a temperature very close to $T_c$. It has been argued that this should 
still be true at finite densities \cite{BB96p}.  
Therefore relativistic mean field models are of particular interest.

Unfortunately until now there is no realistic model of this type which covers
the whole range from low densities up to the chiral phase transition. The
most prominent model in the regime near nuclear matter density is
the Walecka model \cite{SW86} which
describes nuclear binding and the stability of nuclear matter at saturation
density $\rho_o$ quantitatively as the result of the interplay
between a large scalar and a large vector field. Many modifications of
the original Walecka Lagrangian as well as loop corrections have been studied.
However, because of its simplicity and transparency the linear mean field model
(QHD-I) remained the perhaps most important version of the model.
 
Although Walecka theory has been originally designed as an effective theory
for nuclei and nuclear matter at low energies, Walecka-type models are also 
used for the analysis of processes at much higher densities.  For instance
in ref. \cite{LK96p} an extended Walecka model was used to analyze the CERES
data at CERN/SPS \cite{CERES}. Here the initial density is 2.5 $\rho_o$ at
$T$ = 165 MeV. 
It is clear, however, that at some point the Walecka model must become 
unrealistic as there is no chiral phase transition in the model.     
In this situation one might think of extending
the model to negative energy states and to generate the nucleon mass
dynamically. On mean field (Hartree) level this is equivalent to the
Nambu--Jona-Lasinio (NJL) model \cite{NJL} with scalar and vector-isoscalar 
interaction. 
The chiral phase transition in the NJL model has been investigated
by many authors, e.g. \cite{BM87}-\cite{LK92}.     

It was found, however, by Koch et al. \cite{KB87} that in the NJL model, in 
contrast to the Walecka model, there is no saturation density for stable
matter, i.e. the matter described in this model either always expands or
collapses, depending on the vector coupling strength. (A similar result was
obtained by da Provid\^encia et al. \cite{PR87} who investigated the stability
of quark droplets in the NJL model within the time-dependent Hartree-Fock
formalism. They found only expanding droplets.) In order to get the
saturation feature back the authors of ref. \cite{KB87} introduced a new term
to the Lagrangian, which effectively made the scalar coupling constant 
density dependent. 
In section 3 of the present work we reinvestigate this problem using the
formalism of ref. \cite{AY89} which makes more transparent {\it why} the
saturation properties of the Walecka model get lost in a chiral model.
A better understanding of this point may help us to construct an improved model
in the future
We will show that the mechanism which stabilizes matter in the Walecka model
is basically the same as the mechanism which causes a first-order phase
transition in the NJL model, 
namely the existence of two minima in the thermodynamic potential.
In fact, under certain conditions we find stable matter also in the NJL model,
however only in the chirally restored phase.

In section 4 we discuss how the NJL phase transition should be interpreted in
the light of our results.
A general problem of describing the chiral phase transition within the NJL
model is, that one is dealing with the wrong or at least with an incomplete set
of degrees of freedom.  In ref. \cite{ZH94} some improvement was made by going
beyond the standard mean field approach and including mesonic (collective)
degrees of freedom. It is well known that the low-temperature behavior is
dominated by the pion as the lightest particle. However, for the finite
density phase transition -- which is the main task of this paper -- the more
serious question seems to be whether one should deal with nucleons, like in
Walecka theory or the original NJL model or with quarks like in
most NJL papers of the past two decades. If the chiral phase transition comes
together with the confinement-deconfinement one
we should have quarks (and gluons) in one phase and hadrons in the other. 
It is obvious that the hadronic phase is not well described by a gas of
interacting quarks. On the other hand in QCD chiral symmetry is an
(approximate) symmetry for quarks which does not necessarily translate into
a symmetry for nucleons in an effective Lagrangian. Ultimately, 
hadronic matter should be described in terms of hadrons made out of quarks,
perhaps like in the Guichon model \cite{Gu88,ST94} where interacting MIT bags 
are used. In this context our result
that stable quark matter in the NJL model does only exist in the chirally 
restored phase, may be seen as an interesting parallel to bag models where
the hadrons also consist of massless quarks. This will be discussed in more
detail.

\section{Formalism}

\setcounter{equation}{0}

We consider the following generalized NJL-Lagrangian:
\be
{\cal L} \= \psibar (i\gamma^\mu\partial_\mu - m_o) \psi 
            \+ G_S [ (\psibar\psi)^2
             + (\psibar i\gamma_5 \btau \psi)^2] 
            \;-\; G_V (\psibar\gamma^\mu\psi)^2          \;.
\ee
Here $\psi$ is a fermion field with $n_f=2$ flavors and $n_c$ colors. 
It may be interpreted as a nucleon ($n_c=1$) or as a quark field ($n_c=3$). 
Apart from the bare mass $m_o$, the Lagrangian eq.~(2.1) is chirally symmetric
($SU(2)_L \times SU(2)_R$). $G_S$ and $G_V$ are positive constants of dimension
$\rm{mass}^{-2}$, which will be fixed later.

Expanding $\bar\psi\psi$ and $\bar\psi\gamma^\mu\psi$ about their thermal
expectation values we can derive the mean field thermodynamic potential at 
temperature $T$ and chemical potential $\mu$ \cite{AY89}. 
We restrict ourselves to the Hartree approximation. Furthermore in this
paper we only consider $T=0$ and $\mu\geq0$. The result for the thermodynamic
potential is
\be
\omega_{MF}(\mu;m,\mu_R)
\= \omega_m^{(vac)} \+ \omega_m^{(med)}(\mu_R)   
\+ \frac{(m-m_o)^2}{4G_S} \;-\;\frac{(\mu-\mu_R)^2}{4G_V} \;,
\ee
with
\be
\omega_m^{(vac)} \= -(2n_f n_c) \idp E_p 
\ee
and 
\be
\omega_m^{(med)}(\mu_R)
 \= -(2n_f n_c)\idp \;(\mu_R-E_p)\;\theta(p_F - p)  \;,
\ee
being the vacuum part and the medium part of the thermodynamic potential 
(per volume) of a free fermion with mass $m$. $E_p = \sqrt{m^2 + {\bf p}^2}$ 
is its on-shell energy at momentum $\bf p$. The Fermi momentum is given by
$p_F \= \sqrt{\mu_R^2 - m^2} \,\theta(\mu_R^2-m^2)$.
The vacuum part (eq.~(2.3)) is strongly divergent and has to be regularized.
For simplicity we use a sharp cut-off $\Lambda$ in three dimensional momentum 
space.

In addition to the external parameter $\mu$, $\omega_{MF}$ depends
on two other parameters, the dynamical fermion mass $m$ and the renormalized 
chemical potential $\mu_R$, which are related to the scalar density $\pbp$ 
and the vector density $\pdp$ at the chemical potential $\mu$ by \cite{AY89}:
\ba
m &\=& m_o \;-\; 2G_S\pbp  \;,
\cr
\mu_R &\=& \mu \;-\;2G_V\pdp \;.  \cr
\ea
These parameters have to be determined selfconsistently by
calculating $\pbp$ and $\pdp$ from $\omega_{MF}$. It can be shown that 
the selfconsistent solutions correspond to the extrema of
$\omega_{MF}$ as a function of $m$ and $\mu_R$. 
This leads to a set of coupled equations for $m$ and $\mu_R$ which reads for
$T=0$ and $\mu\geq 0$:
\be
m \= m_o \+ 2G_S \; (2n_f n_c) (\idp \; \frac{m}{E_p}
\;-\; \idp \; \frac{m}{E_p}
\;\theta(p_F-p)) \;,
\ee
\be
\mu_R \= \mu  \;-\; 2G_V\,(2n_f n_c) \idp 
\;\theta(p_F-p) \;. \hskip3.5cm
\ee
Eq.~(2.6) is the well-known NJL gap equation for the
dynamical mass $m$ at temperature $T=0$ and chemical potential $\mu\geq 0$.
In general this equation has more than one solution and the physical one is
the one which minimizes $\omega_{MF}$. It is therefore more safe to start the
analysis from the thermodynamic potential (eq.~(2.2)) than from the gap
equation (eq.~(2.6)) because the latter does not tell us which solution
is the correct one. However, eq.~(2.7) has a unique solution for
$\mu_R$ and we can use this equation to eliminate the $\mu_R$ dependence of
$\omega_{MF}$. Thus for a given $\mu$ we are left with a function which
depends on $m$ only. More precisely, we define:
\be
\tilde\omega(\mu;m) \= \omega_{MF}(\mu;m,\mu_R(\mu,m)) -
\omega_{MF}(0,0,m_{vac},0) \;.
\ee
Here $\mu_R(\mu,m)$ is the solution of eq.~(2.7) for given $\mu$ and
$m$, while $m_{vac}$ is the dynamical mass which minimizes the thermodynamic
potential of the vacuum. 
So the second term on the r.h.s. is just a constant which shifts the minimum
of the vacuum thermodynamic potential to zero.
Since the absolute magnitude of the thermodynamic potential has no physical
meaning, we can always subtract such a constant.

Once we have the thermodynamic potential, other thermodynamic quantities can
be calculated in the standard way. In particular we have for the baryon number
density, energy density and pressure
\be
\rho_B \= \frac{1}{n_c} \, \pdp \= \frac{n_f}{3\pi^2}\,p_F^3 \;,\quad
\varepsilon \= \tilde\omega \+ \mu\,n_c\,\rho_B \;,\quad
p = - \tilde\omega \;,
\ee
where the energy and the pressure of the physical vacuum was defined to be zero. 
Then for a given
$\mu$ these formulas have to be evaluated for the values of $m$
and $\mu_R$ which minimize the thermodynamic potential.

For comparison we also consider the QHD-I version
of the Walecka model \cite{SW86} in classical mean field approximation. 
The Lagrangian of this model reads:
\ba
{\cal L}_W &\=& \psibar [\gamma^\mu(i\partial_\mu - g_v\omega_\mu)
                       - (m_N - g_s\sigma)] \psi \cr
           & &\+ \frac{1}{2} (\partial_\mu\sigma\partial^\mu\sigma -
               m_s^2\sigma^2) 
           \;-\; \frac{1}{4} F_{\mu\nu} F^{\mu\nu} \+ 
           \frac{1}{2} m_v^2 \,\omega_\mu\omega^\mu  \;,\cr 
\ea
with a nucleon field $\psi$, a scalar field $\sigma$ and a vector field
$\omega$. 
The classical mean field theory which results from solving the equations of
motion after replacing the boson fields by their space-time independent
expectation values is equivalent to the NJL model in mean field approximation
with $n_c=1$ and $m_o=m_N$ (the free nucleon mass), without taking into account
negative energy states, i.e. neglecting the vacuum part of the free
thermodynamic potential (eq.~(2.3)). 
Here we made the identification
\be
G_S \= \frac{1}{2}\,\frac{g_s^2}{m_s^2} \;,\hskip2cm
G_V \= \frac{1}{2}\,\frac{g_v^2}{m_v^2} \;.
\ee
So, keeping the NJL notation, the Walecka thermodynamic 
potential can be written:
\be
\omega_W(\mu;m,\mu_R)  
\= \omega_m^{med}(\mu_R)
      \+ \frac{(m-m_N)^2}{4G_S} \;-\;\frac{(\mu-\mu_R)^2}{4G_V} \;,
\ee
and all thermodynamic properties can be derived in the analogous way
as discussed above for the NJL model.

\section{Stable Matter Solutions}

\setcounter{equation}{0}

The saturation of nuclear matter in the Walecka mean field model is usually
discussed in terms of the energy per nucleon which (with the appropriate choice
of the model parameters) has a minimum at nuclear matter density $\rho_o$.
In a similar analysis Koch et al. \cite{KB87} found for the NJL mean field
that there is no such minimum in the energy per nucleon and consequently
no stable matter. In order to understand how this difference comes about it is
useful to recall that the stable nuclear matter in the Walecka model is related
to a first-order gas-liquid phase-transition with the nuclear matter belonging
to the liquid phase: If the average density of nucleons is less than $\rho_o$
the nucleons are not uniformly distributed but a mixed phase is formed 
consisting
of nucleon droplets of density $\rho_B = \rho_o$ surrounded by the vacuum with 
$\rho_B = 0$ (at $T = 0$). Now Gibbs' criteria tell us that the pressure $p^*$
and the chemical potential $\mu^*$ in the droplets must be equal to the pressure 
and the chemical potential in the vacuum, i.e. $p^*=0$ and $\mu^*<m_{vac}$. 
In other words: At some chemical potential $\mu^*<m_{vac}$ the thermodynamic 
potential must have an {\it additional minimum} at $m=m^*<m_{vac}$ with 
\be
\tilde\omega(\mu^*;m^*) \= 
\tilde\omega(\mu^*;m_{vac}) \= 0 \;; \hskip5mm |m^*| < \mu^* < m_{vac} \;.  
\ee
The existence of this additional minimum is the condition for the existence of 
stable matter. Obviously $\mu^*$ is just the critical chemical potential for 
the gas-liquid phase transition.

In order to illustrate this point, in the left panel of fig.~1 the Walecka 
thermodynamic potential is plotted as a function of the mass parameter $m$ for 
four different chemical potentials. The solid line corresponds to the critical 
chemical potential $\mu^*$ where eq.~(3.1) is satisfied by a minimum at an 
effective nucleon mass $m^*=$ 514.5 MeV.

As discussed in the previous section the essential difference between the
Walecka and the NJL mean field is the existence of the vacuum term 
$\omega_m^{(vac)}$ (eq.~(2.3)) in the latter. It is this term which - for
a sufficiently large scalar coupling - causes the spontaneous breaking of
chiral symmetry in the NJL model, i.e. a finite vacuum mass $m_{vac}$ while
the bare mass $m_o$ is zero or at least small.
At some critical chemical potential the system undergoes a  phase transition
and chiral symmetry gets restored. This chiral phase transition is the analogue
to the gas-liquid phase transition of the Walecka model: The chirally restored
phase corresponds to the liquid and the broken phase to the gas. So, according
to our previous discussion we expect the possibility of stable matter in the
NJL model but only in the chirally restored phase.  

As an example we consider quarks in the chiral limit.
Using standard methods \cite{Kle} we can more or less fix the scalar coupling
constant $G_S$ and the cut-off $\Lambda$ by fitting the pion decay constant
$f_\pi$ and the quark condensate $\ave{\bar u u}$ in the vacuum. Since the
latter is not known very accurately we investigate three sets of parameters
with different constituent quark masses. They are listed in table~1. 
We do not fix the vector coupling constant $G_V$ but treat it as a free
parameter and discuss its influence on the results. 
 
An important quantity for our later discussion is the bag constant
\be
B \= \tilde\omega(\mu=0; m=0) \;,
\ee
i.e. the pressure difference between the physical vacuum ($m =m_{vac}$)
and the trivial vacuum ($m=0$). Its values for the three different parameter
sets are also given in table~1.
 
We start our discussion with parameter set 2 and $G_V=0$. The corresponding
thermodynamic potential $\tilde\omega$ is plotted in the right panel of fig.~1
for various chemical potentials. In the chiral limit $\tilde\omega$ is symmetric
in $m$. We show only the positive mass region.
The dotted line corresponds to the vacuum thermodynamic potential ($\mu$ = 0).
The potential has a minimum at $m =$ 400 MeV (the vacuum mass we have chosen) 
and a maximum at $m$ = 0.
At $\mu =$ 344.4 MeV the $m = 0$ maximum turns into
a minimum, which becomes zero at
$\mu^* =$ 378.5 MeV. Finally at $\mu = 409.3$ MeV the original minimum 
disappears. So, according to eq.~(3.1) we expect stable quark 
matter at $\mu^* =$ 378.5 MeV with massless quarks. This corresponds
to a baryon number density of 2.8 times nuclear matter density.
This is confirmed by the behavior of the energy per baryon number 
($\varepsilon/\rho_B$) as a function of the baryon number density which is
shown in fig.~2. The solid line corresponds to the massive 
solutions of the gap equation, the dashed line to the massless ones. 
The energetically favored point is
$\rho_B = 2.8 \rho_o$ and lies in the chirally restored regime, as we expected.

The phase transition in this example is of first order and takes place at zero
density. In fact, according to eq.~(3.1) it has to be of this type in order
to allow for stable matter solutions. However, the character of the phase
transition depends on the model parameters, in particular on the vector 
coupling constant: For parameter set 2 and $0.3 G_S < G_V < 0.8 G_S$ there is
a first-order phase transition but with a finite transition density (i.e.
$\mu^*>m_{vac}$), for $G_V > 0.8 G_S$ the phase transition is of second order. 
In both cases there is no stable matter solution. 
This can be seen from fig.~3 where again the
energy per baryon number is plotted as a function of density. The left panel
corresponds to $G_V=\frac{1}{2} G_S$, the right one to $G_V=G_S$. Obviously the
energetically favored point is $\rho_B = 0$ in both cases, i.e. the matter is
unstable against expansion. 
 
Our analysis and in particular the comparison between the left and the right
panel of fig.~1 give some insight why the observation of Koch et al. \cite{KB87}   
that the saturation properties of the Walecka model get lost when one tries to
make it chirally symmetric, was not an accidental coincidence. 
These authors only considered the massive solutions of the gap equation and
consequently did not find any stable matter in the NJL model. The fact, that
saturated nuclear matter with massive nucleons does exist in the Walecka model,
is a direct consequence of its explicitly broken symmetry: The $(m-m_N)^2$
term in eq.~(2.12) interferes with the $m$-symmetric $\omega_m^{med}$, leading
to a minimum somewhere in between $m=0$ and $m=m_N$. In the chirally symmetric
model (NJL in the chiral limit) this minimum still exists but is shifted to
the symmetric point $m=0$ and is identical to the minimum which is
responsible for the chiral symmetry restoration at finite densities.  

For parameter set 3 we find qualitatively similar results as for set 2:
For $G_V < 0.6 G_S$ there is a first-order phase transition with zero transition
density which leads to stable droplets of massless fermions, whereas there is
no stable matter for $G_V > 0.6 G_S$. For parameter set 1
the transition density is always greater than zero for positive
$G_V$ and there is no stable matter solution. 

The main properties of the stable matter solutions are shown in fig.~4 as a
function of the vector coupling constant $G_V$. The saturation density $\rho_S$
is plotted in the upper panel, while the lower panel shows the binding energy
per quark
\be
E_B \= - (\frac{\varepsilon}{n_c\rho_S} \;-\; m_{vac}) \;.
\ee
The dashed lines correspond to parameter set 2, the solid lines to set 3.
With increasing vector coupling the binding energy drops almost linearly.
As a result of the increasing repulsion
the saturation density also drops with $G_V$, but more slowly. Because of the
larger bag constant the stable matter is denser and bound more tightly for
parameter set 3 than for parameter set 2. This is also the reason why we did
not find any stable matter for parameter set 1, where the bag constant was
very small.

Our results can be understood almost analytically. For $m = 0$ the
energy per baryon number is given by
\be
\frac{\varepsilon(\rho_B)}{\rho_B} \= \frac{B}{\rho_B} \+ \frac{3 n_c}{4}
\left(\frac{3\pi^2}{n_f} \right)^{1/3}\,\rho_B^{1/3} \+ G_V n_c^2 \rho_B \,.
\ee
Minimizing this formula we find for the saturation density $\rho_S$
\be
\frac{n_c}{4} \, \left(\frac{3\pi^2}{n_f} \right)^{1/3}\,\rho_S^{4/3}
\+ G_V n_c^2 \rho_S^2 \= B \,,
\ee
provided the energy per baryon number at this point is less than $n_c m_{vac}$,
the value at $\rho_B = 0$. So for $G_V = 0$ $\rho_S$ behaves like $B^{3/4}$
and it gets reduced by a finite vector interaction. The energy density at this
point is
\be
\varepsilon(\rho_S) \= 4B \;-\; 2G_V n_c^2 \rho_S^2     \;.
\ee
Solving eq.~(3.5) for $\rho_S$ until linear order in $G_V$ and inserting the
result into eq.~(3.6) we finally get for
the binding energy per quark
\be
E_B \;\simeq\; m_{vac} \;-\;
\left(\frac{12\pi^2}{n_f n_c} B \right)^{1/4} \; [ 1 + 2G_V 
(\frac{n_f n_c}{3\pi^2} B)^{1/2} ] \;,
\ee
where we neglected terms of order $G_V^2$.

For $\Lambda^2 \gg m_{vac}^2$ the following approximate relation holds 
\cite{AB}:
\be
B \;\simeq\; \frac{1}{2}\,m_{vac}^2\,f_\pi^2 \+ 
\frac{n_fn_c}{32\pi^2}\,m_{vac}^4 \;.
\ee
Although the cut-off is not very large compared to $m_{vac}$ for our examples,
this formula works better than 10\% for parameter set 1 and even for set 3,
where $\Lambda$ and $m_{vac}$ are almost equal, the error is only 30\%.
Inserting this formula into eq.~(3.7) and taking $G_V=0$ we can estimate the
minimal value of $m_{vac}$ for which $E_B > 0$ is possible. The result is
\be 
   m_{vac} \;\gsim\; \left(\frac{48}{5n_fn_c}\right)^{1/2} \,\pi\,f_\pi
   \;\simeq\; 4f_\pi  \;,
\ee
where we have evaluated the expression for $n_f = 2$ and $n_c = 3$. This
result is consistent with the fact that we found stable matter solutions for 
parameter set~2 but not for set~1. It is quite remarkable that 
$g = m_{vac}/f_\pi \gsim 4$ was also found in ref. \cite{DB92} as the critical
condition for solitonic solutions in the NJL model if a sharp O(3) cut-off is 
used. Reinserting eq.~(3.9) into eq.~(3.8) we find a minimal bag constant 
for stable matter solutions $B \gsim$ (175 MeV)$^4 \simeq$ 125 MeV fm$^{-3}$,
about twice the MIT bag constant \cite{DG75}.

\section{Discussion and Conclusions}

\setcounter{equation}{0}

In this section we want to discuss what the results of section 3 mean for
the interpretation of the NJL phase transition. First let us come 
back to the question whether nucleon or quark degrees of freedom should be 
used in the model. Starting from low densities, nucleons
seem to be the more natural choice. However, as outlined above, such a 
model would already fail to describe ordinary nuclear matter. Here the fact
that we found a stable matter solution in the chirally restored phase does
not help us. To the extent finite size effects can be neglected, this would 
mean that nuclei consist of massless nucleons, a picture which is certainly
unrealistic. If we introduce a small current mass $m_o$, we could obtain a
finite effective mass, but still $m^*$ would
stay well below the Walecka effective nucleon mass which is already considered
to be too small. So there is no simple extension of the Walecka model which
keeps its low-density behavior but also allows for a chiral phase transition.
A possible way out would be to add extra terms to the Lagrangian eq.~(2.1)
as proposed in ref. \cite{KB87}.

The alternative option is to interpret the NJL model as a quark model, as it 
is done by most present authors. Although it is clear that quarks are not the 
correct
degrees of freedom in the hadronic phase, such models agree reasonably well
with model independent low-density theorems if one uses simple translation
rules between nucleon and quark quantities (e.g. $m_Q = m_N/3$, $\sigma_{\pi Q}
= \sigma_{\pi N}/3$ etc.) \cite{LK92}.  The hope behind this is that - 
concerning the phase transition - the quarks behave more or less like free, 
even though they are confined in nucleons. On the other hand the stability 
problem is usually taken less serious than in models with nucleon degrees of 
freedom, since there is no such thing like stable quark matter anyway in 
nature. This point of view is of course very questionable.

If we assume that the chiral phase transition for quarks takes place at a
finite density (i.e. a second-order phase transition or a first-order phase
transition with a non-zero transition density) our analysis shows that the
quarks, if they really behave like quarks in an NJL mean field, should repel
each other and the matter should tend to expand. However, the confinement
prevents this expansion, at least for individual quarks. So as long as the
confinement radius is smaller than the average distance between the quarks,
i.e. as  long as the nucleons do not overlap, the confining forces (or other
binding forces inside the nucleons) cannot be negligible compared to the
scalar and vector mean fields.

These binding forces are provided by the mean fields themselves if we assume a
first-order phase transition with zero transition density on the quark level. 
In this case
we would find droplets of massless quarks and the density is $\rho_B=\rho_S$ 
inside these droplets and $\rho_B=0$ outside. This mixed phase could be
interpreted as a schematic picture of a bag model description. In
a bag model we also have massless quarks at a relatively high density inside
the bag and zero density outside. 

In fact, in the MIT bag model \cite{MIT} with massless non-interacting quarks
the ground state energy of a non-strange baryon is given by \cite{DG75}
\be
E_{MIT} \= \frac{4\pi}{3} B R^3 \+ \frac{3x-z_o}{R}
\= \frac{B}{\rho_B} \+ (3x-z_o)\,\left(\frac{4\pi}{3}\right)^{1/3}\,
\rho_B^{1/3} \;,
\ee
where $R=(\frac{3}{4\pi\rho_B})^{1/3}$ is the bag radius, $x=2.04$ and $z_o$
a constant which accounts for the zero point motion. So the structure of 
$E_{MIT}$ is exactly the same as that of  $\varepsilon/\rho_B$ for massless 
quarks
in the NJL model (eq.~(3.4)) if we switch off the vector interaction. This is
of course not very surprising. Switching off the vector interaction there is
no mean field left in the chirally restored phase and we are dealing with a
gas of free massless quarks stabilized only by the bag pressure. So our
stable droplets are nothing but large quark bags. The difference between
eq.~(3.4) and eq.~(4.1) is that in our thermodynamic approach we assumed a
very large number of quarks inside a large volume whereas eq.~(4.1) was
derived using the correct boundary conditions for three quarks inside a 
volume which can be small. This leads to a different coefficient of the
$\rho_B^{1/3}$ term which dominates the energy for small radii. However,
this difference turns out not to be terribly large. For $z_o=2.69$ it
would even vanish. For $z_o=1.84$ as in ref. \cite{DG75} the coefficients of
$\rho_B^{1/3}$ differ by about 20\%. This corresponds to a 5\% difference
in the bag radii. As discussed in section 3 the original MIT bag constant is too
small to produce stable matter in the NJL-model. The stability densities
we find with the larger bag constants of the parameter sets 2 and 3 (see
fig.~4) correspond to bag radii of 0.7-0.8 fm. 

In order to calculate baryon masses the bag energy eq.~(4.1) should be
corrected for spurious center of momentum motion \cite{cm}. Again, these
correction terms are not present in our thermodynamic approach because they
are suppressed if the number of particles inside the droplet is large. For
3-quark bags, however, they reduce the radii by 20-30\%, which means the
density becomes drastically enhanced.

Usually a bag model picture of a baryon is considered to be quite different
from a constituent quark model. However, as just discussed, the NJL mean
field may lead to the formation of bag-like quark droplets while there are
constituent quarks in the model as well. Moreover, since the binding energies
we found in the previous section are less than 15\% of $m_{vac}$ (see fig.~4)
the bag energies are roughly given by the constituent quark relation
$\varepsilon/\rho_B \simeq 3 m_{vac}$. The essential difference to the 
constituent quark model is that in our picture the constituent quark mass
is {\it not} the mass of the quarks inside a baryon but the mass of a single
quark state in the vacuum.\footnote{On the other hand baryons in the NJL model
can also be described by solving Fadeev equations for three interacting 
constituent quarks \cite{Fad}.
In the soliton description $\bar\psi\psi$ strongly varies inside the
baryon and even changes sign at small radii (see e.g. \cite{sol}).  So
in this model the constituent quark mass is again the mass of a single quark 
in the vacuum rather than in a baryon.
It is possible that the question about the quark mass inside a baryon is more
a question of the most efficient way to sum diagrams (which could be different
for different processes under consideration) than a meaningful physical
question.}

These constituent quark states, which do not exist in the bag model are also
the reason why the NJL model does not confine. For $\rho_B\rightarrow\infty$
the energy per baryon behaves like $B/\rho_B$ for both, the MIT bag as well
as for droplets consisting of the massless solutions of the NJL gap equation.
So in order to increase the MIT bag radius to infinity an infinite amount of
energy is needed. The same would be true for the NJL droplets if only the
trivial solutions existed. However, when the density gets low enough the
quarks can acquire a constituent mass and only a finite amount of energy is
needed to increase the radius to infinity. So amusingly the same mechanism
which generates the bag pressure, namely spontaneous symmetry
breaking, prevents it from confining the quarks. In the MIT bag model, where 
the bag pressure is put in by hand, this does not happen.  
 
Another difference between the NJL model and the bag model shows up when we
go away from the chiral limit and introduce a small current mass $m_o$.
Then the bag model quarks simply have this mass $m_o$. In the NJL droplets,
although still much smaller than $m_{vac}$, $m^*$  
is considerably larger than $m_o$. For $m_o = 5.5$ MeV 
this is shown in fig.~5.
For $G_V = 0$ $m^*$ is about 40 MeV for parameter set 2 and about 
20 MeV for parameter set 3 and it grows for both sets with increasing $G_V$.
Qualitatively this behavior follows from the fact that for $m_o \neq 0$ the
solutions $m(\rho_B)$ of the gap equation do not show a sharp transition to
$m = m_o$ but have a long smooth tail. Since the saturation
density $\rho_S$ is a decreasing function of $G_V$ the corresponding masses
$m^* = m(\rho_S)$ increase with $G_V$.  

In the NJL model the constituent quark mass is closely related to the quark
condensate in the vacuum (eq.~(2.5)) which is the physically more meaningful
quantity. Inside the baryon droplets the quark condensate is about an order of
magnitude lower for $m_o=5.5$ MeV and even zero in the chiral limit. The
baryons are good ``vacuum cleaners''. If we integrate the difference between
the quark condensate in the bag and in the vacuum over the bag volume we get
\be
\bra{B} \bar uu + \bar dd \ket{B} \= \frac{1}{\rho_S}
\,(\,\pbp|_{\rho_B=\rho_S} \,-\, \pbp|_{\rho_B=0})  \;.
\ee
Here ``$B$'' stands for ``baryon'' or ``bag''. In the chiral limit the r.h.s.
reduces to  $-\ave{\bar\psi\psi}/\rho_S$
which is between 8.0 and 9.3 for parameter set 2 and between 5.3 and 7.7 for
parameter set 3, depending on $G_V$. For $m_o=5.5$ MeV we find somewhat smaller
values, between 6.8 and 7.4 for set 2 and between 4.9 and 6.4 for set 3. 
Although the
model is very crude these results are in qualitative agreement with the value
determined for the nucleon from $\pi N$ scattering:
\be 
\bra{N} \bar uu + \bar dd \ket{N} \= \frac{\sigma_{\pi
N}}{\frac{1}{2}(m_u^{(o)}+m_d^{(o)})} \;\simeq\; 8 \;,
\ee
where we have used $\frac{1}{2}(m_u^{(o)}+m_d^{(o)}) = m_o = 5.5$ MeV, as
before, and the $\pi N$ sigma term $\sigma_{\pi N} = 45$ MeV \cite{GL91}.

If the baryon droplets do not further interact, the system will stay in the
vacuum-droplet mixed phase until the average density reaches the stability
density $\rho_B=\rho_S$. Above this density there is no empty space left and
the total volume is uniformly occupied by massless quarks.
This point defines the ``macroscopic'' finite density phase transition
compared to the ``microscopic'' phase transition at zero density which was the
starting point of this discussion. So the ``macroscopic'' phase transition
would take place at $\rho_B \sim \rho_S$ which can be substantially higher
than the critical densities the NJL model predicts even for second order phase
transitions. For quark degrees of freedom the latter are unrealistically low.

Whereas we could interpret the quark droplets of our thermodynamic approach
as schematic baryon bags, there is nothing which would correspond to mesonic
bags in our picture. Mesons have zero baryon number and therefore cannot
produce a chirally restored bag in their surrounding. However, mesons do of
course exist in the NJL model as collective quark-antiquark excitations of the
vacuum
or the medium. In fact, this picture is much more powerful than the bag model
picture, since it leads to pions as massless Goldstone bosons, whereas the
MIT bag pion came out too heavy \cite{DG75}. 

The NJL mesons could mediate interactions between the
baryon-droplets. The resulting picture is very similar to the Guichon model
\cite{Gu88,ST94} where MIT bags interact by the exchange of scalar and vector 
mesons. In this way one might obtain a Walecka-like description of nuclear 
matter. Our schematic picture shows a way,
how the two ingredients of the Guichon model, the bag pressure and the
mesonic interaction could be traced back to the same origin.

The main problem to construct such a model quantitatively is the question how
to couple the mesons in the nontrivial vacuum to the quarks in the bag. The
meson properties are quite different inside and outside the bag and one has
to take into account boundary effects in an appropriate way. This would be
the main difference to the Guichon model where the mesons do not change their
properties at the bag surface. 
Here a better
description of the droplet mixed phase which goes beyond the assumption of
constant mean fields seems to be necessary. 

In summary, we reinvestigated the question of matter stability within 
the Nambu--Jona-Lasinio model in mean field approximation. We confirmed 
earlier results that the NJL model does not allow for stable quark or nuclear 
matter in the phase with spontaneously broken chiral symmetry but we 
found stable matter solutions in the chirally restored phase. 
In the model this gives rise to
the formation of quark droplets which are stabilized by the bag pressure of the
surrounding vacuum. These droplets show many similarities to bag model
solutions. Although we failed to describe {\it hadronic} matter within the 
NJL model we argue that this might be 
achieved in a second step by considering meson 
exchange between the bags, similar to the Guichon model.

\vskip 1cm      
I would like to thank G.E. Brown for useful discussions.

\newpage
\begin{table}
\begin{center}
\begin{tabular}{cccccc}
\hline
& \qquad set 1 \qquad & \qquad set 2 \qquad & \qquad set 3 \qquad \\
\hline
$\Lambda$ [MeV] & \qquad 650 & \qquad 600 & \qquad 570 \\
$G_S\Lambda^2$ & \qquad 2.14 & \qquad 2.45 & \qquad 2.84 \\
$m_{vac}$ [MeV] & \qquad 313 & \qquad 400 & \qquad 500 \\
$f_\pi$ [MeV] & \qquad 92.6 & \qquad 93.9 & \qquad 92.5 \\
$\ave{\bar uu}^{1/3}$ [MeV] & \qquad -248.9 & \qquad -244.9 & \qquad -242.6 \\
B [MeV/fm$^3$] \qquad & \qquad 73.9 & \qquad 135.3 & \qquad 224.8 \\
\hline
\end{tabular}
\end{center}
\caption{Three sets of model parameters ($O(3)$ cut-off $\Lambda$ and
scalar coupling constant $G_S$) and the corresponding vacuum properties
for the NJL model with quark degrees of freedom in the chiral limit:
constituent quark mass $m_{vac}$, the pion decay constant $f_\pi$, the
quark condensate $\ave{\bar uu}$ and the bag constant $B$.}
\end{table}


\phantom{empty}
\pagebreak
\pagestyle{empty}
\begin{center}
\centering{\bf{\large Figure Captions}}
\end{center}
\vspace{1.3cm}
\begin{itemize}

\item [{\bf Fig.~1}]
Thermodynamic potential $\tilde\omega$ as a function
of the mass parameter $m$ for different chemical potentials. 

Left panel: Walecka model with $m_N = 939$ MeV, $m_s = 550$ MeV, $m_v = 783$
MeV, $g_s = 10.3$ and $g_v = 12.7$. The different lines correspond to $\mu = 0$
(dotted line), $\mu = 915.1$ MeV (dashed line), $\mu = \mu^* = 923.15$ MeV 
(solid line) and $\mu = 939.65$ MeV (dashed-dotted line).

Right panel: NJL for parameter set 2 (see table 1) and $G_V = 0$. Here the
different lines correspond to $\mu = 0$ (dotted
line), $\mu = 350$ MeV (dashed line), $\mu = \mu^* = 378.5$ MeV (solid line) 
and $\mu = 410$ MeV (dashed-dotted line). 

\item [{\bf Fig.~2}]
Energy per baryon number  $\varepsilon/\rho_B$ as a function of the density 
$\rho_B$ for parameter set 2 and $G_V=0$. The solid line corresponds to the
massive solutions of the gap equation, the dotted line to the massless ones.

\item [{\bf Fig.~3}] 
The same as fig.~2 but for $G_V = \frac{1}{2}G_S$ (left panel) and $G_V =
G_S$ (right panel).
  
\item [{\bf Fig.~4}] 
Saturation density $\rho_S$ (upper panel) and binding energy $E_B$ (lower
panel) of stable quark matter as a function of $G_V/G_S$. The dashed lines 
correspond to parameter set 2, the solid lines to set 3. There is no stable 
quark matter for parameter set 1 in the positive $G_V$ regime.

\item [{\bf Fig.~5}] 
Effective quark masses $m^*$ for stable quark matter as a function of
$G_V/G_S$. The dashed (solid) line corresponds to the scalar
coupling constant $G_S$ and the cut-off $\Lambda$ of parameter set 2 (3), but
using a finite current quark mass $m_o =$ 5.5 MeV.

\end{itemize}

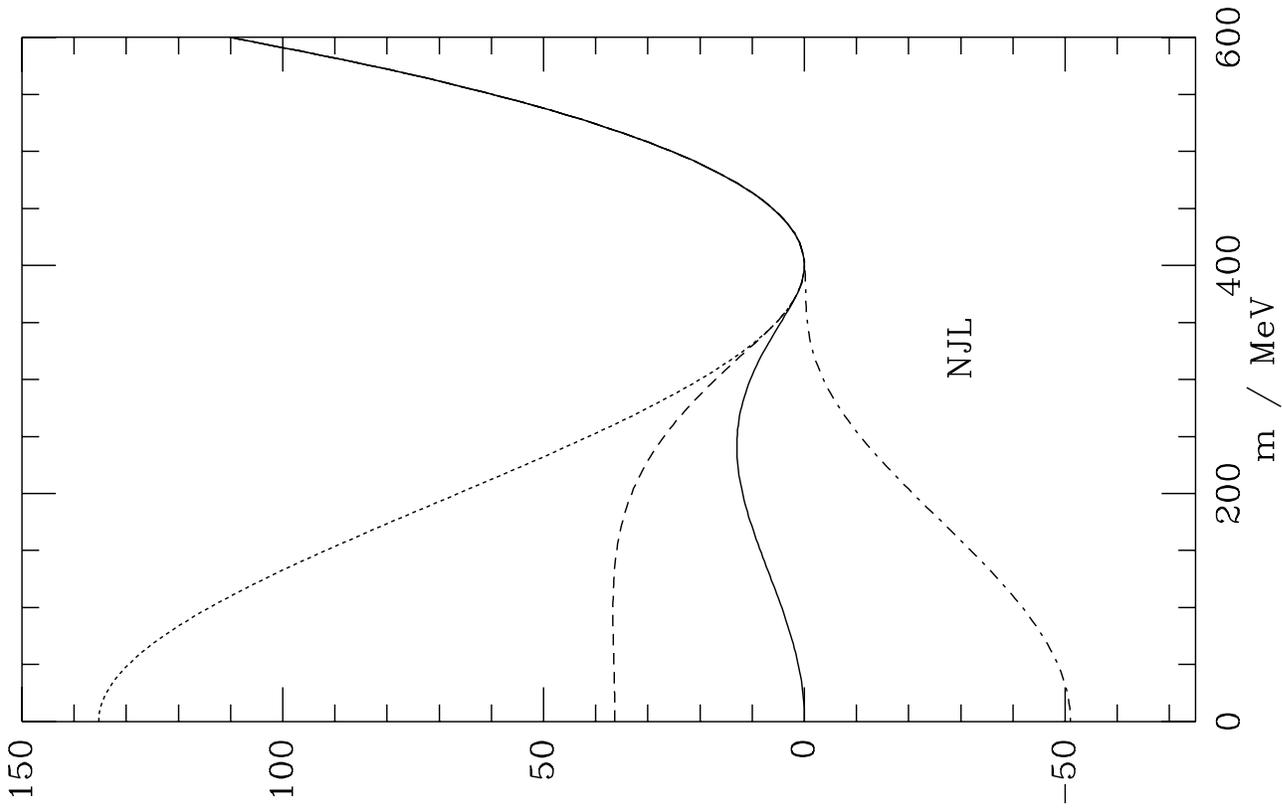
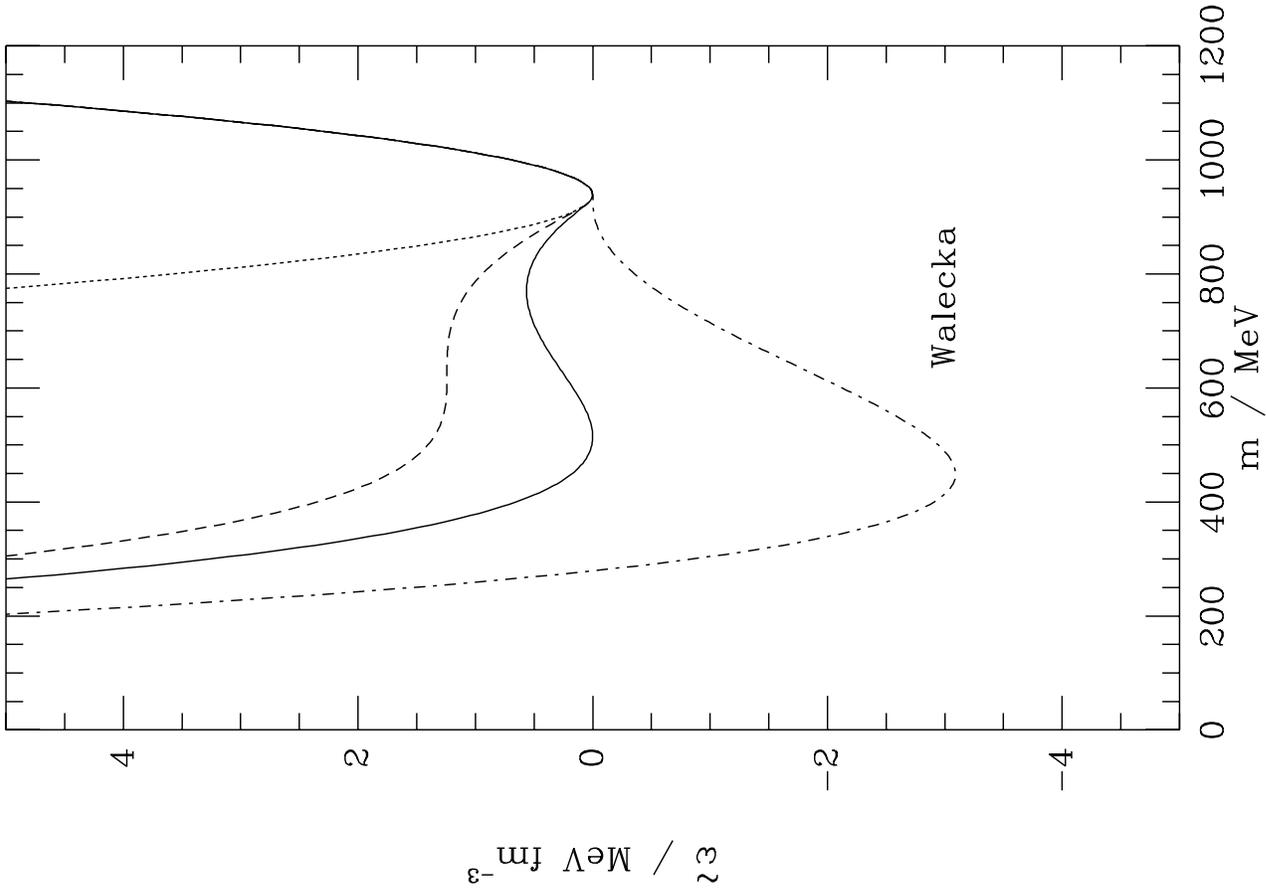
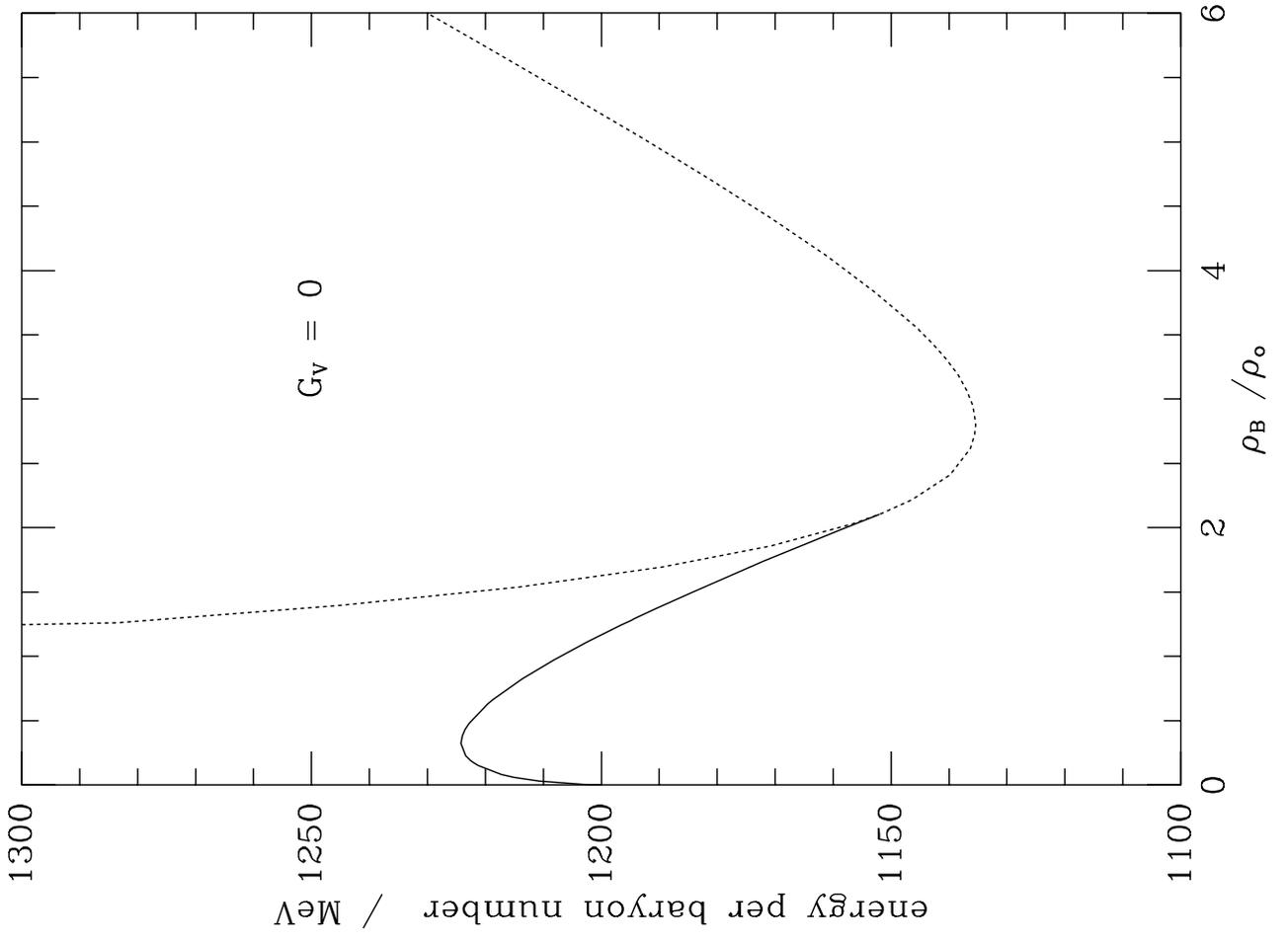
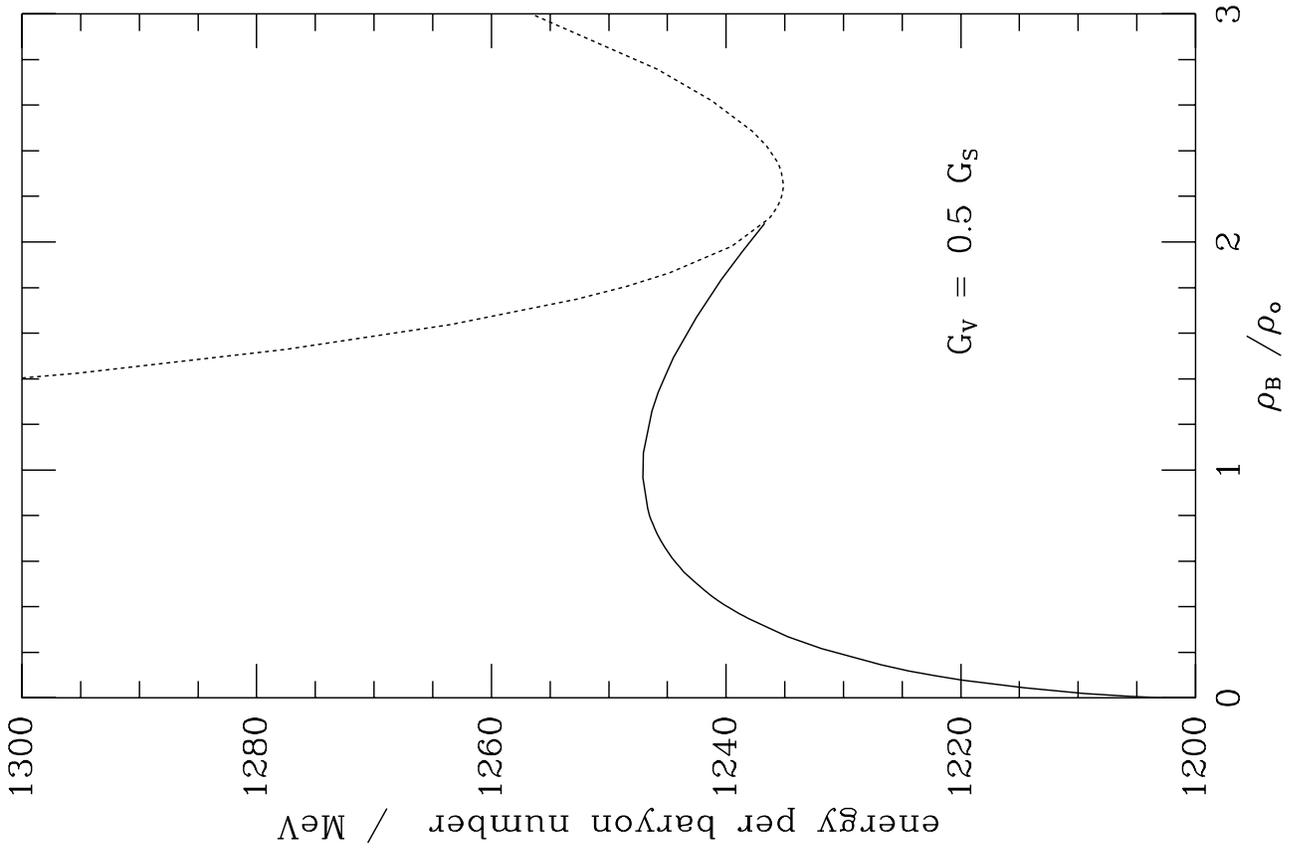
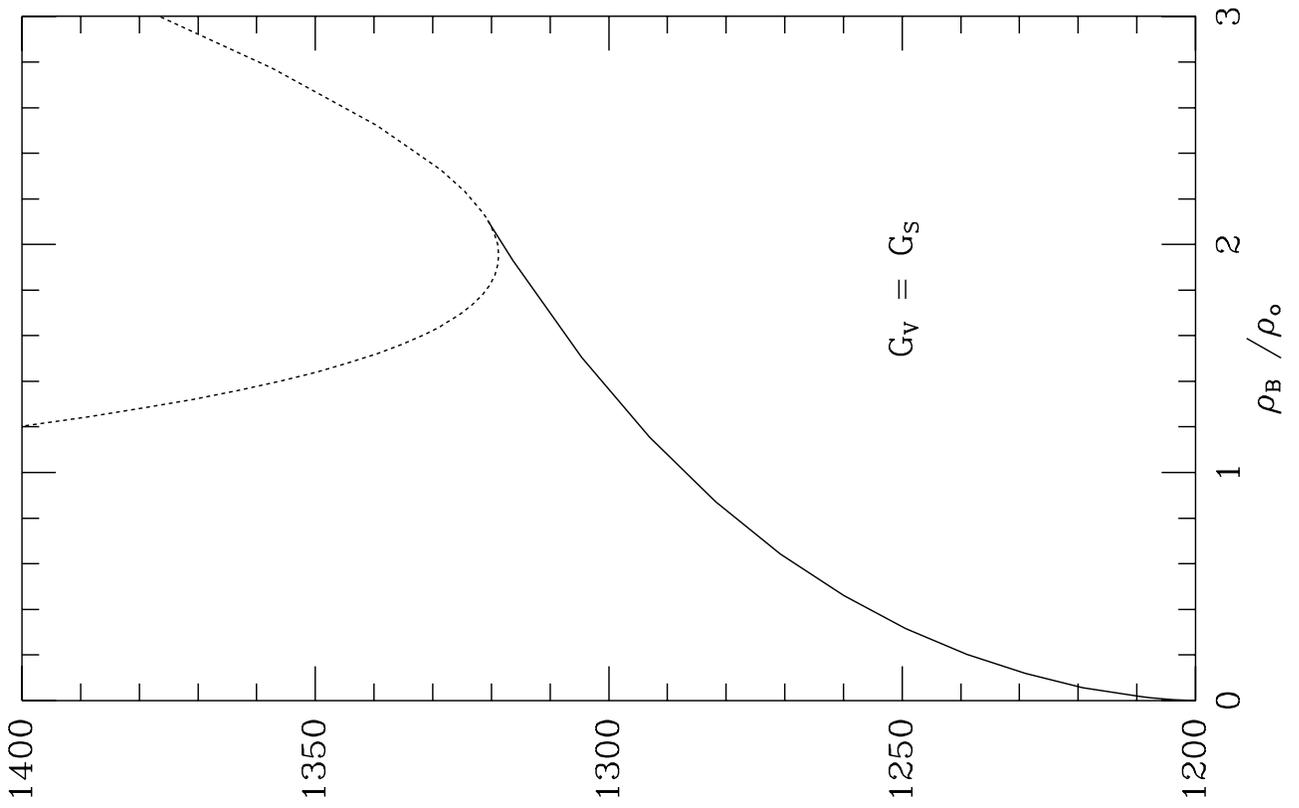
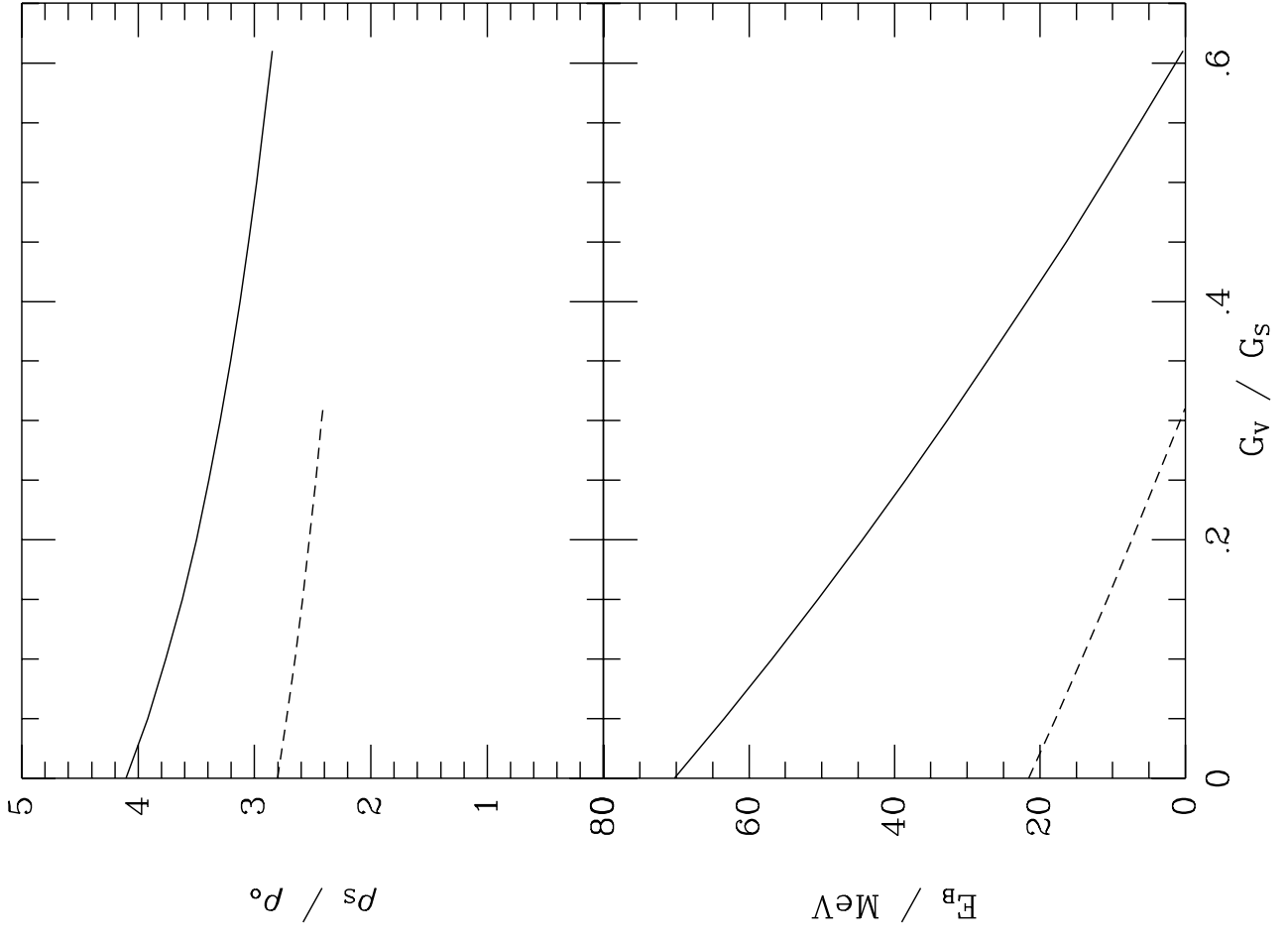
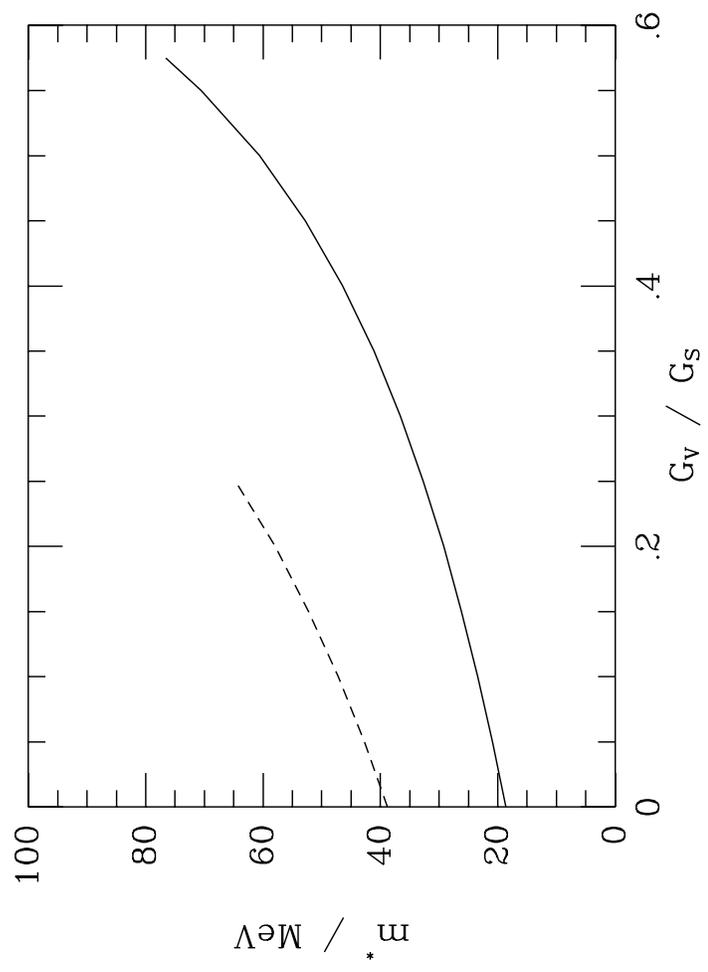
\end{document}